# Systems, Actors and Agents:
## Operation in a multicomponent environment

Mark Burgin

University of California, Los Angeles
405 Hilgard Ave.
Los Angeles, CA 90095

**Abstract**. In this paper, we further develop multi-agent approach by creating new types of system models. The problem is that conventional models of multi-agent and multicomponent systems implicitly or explicitly assume existence of the absolute time or even do not include time in the set of defining parameters. However, it is rationalized theoretically and validated experimentally that there are different times and time scales in a variety of real systems – physical, chemical, biological, social, etc. Thus, the goal of this paper is construction of multi-agent and multicomponent system models with concurrency of processes and diversity of actions. To achieve this goal, a mathematically based system actor model is elaborated and its properties are studied.

## 1. Introduction

Multi-agent approach is becoming more and more popular in the area of computing, networking, artificial intelligence, robotics, distributed control, resource management, collaborative decision support systems, data mining, etc. (Buşoniu, et al, 2010; Shoham and Leyton-Brown, 2008; Vlassis, 2007; Weiss, 1999). Usually, it is assumed that a multi-agent system is a group of autonomous, interacting entities sharing a common environment, which they perceive with sensors and upon which they act with actuators.

Time is a critically important characteristic of any real-life system. However, not all features of time are adequately presented in the conventional multi-agent models.

If we analyze existing approaches and directions in the multi-agent approach, we can see that in all dynamic models of multi-agent systems, either time is implicitly induced by actions of agents and system states or it is explicitly assumed that unique time exists for the whole system. An archetypal example of this situation is the absolute Newtonian time in the physical universe, which is innate for the entire classical physics.

However, relativity theory and various experiments disproved this assumption bringing forth the concept of local time (cf., for example, (Einstein, et al, 1923)). The system theory of time extends this principle much further (Burgin, 1992; 2002). Other researchers also advocated existence of different times or different time scales in their theories (cf., for example, (Prigogine, 1980; Barwise and Seligman, 1997)). Besides, as Norbert Wiener (1961) writes, one of the most famous philosophers of the 20$^{th}$ century Bergson lays special emphasis on the distinction between the reversible time of physics, in which nothing new happens, and the irreversible time of evolution and biology, in which there is always something new (Bergson, 1910). In spite of this, time in general systems theory is similar to time in classical physics, namely, either all models of systems in general systems theory are still based on the principle of absolute (global) time or time is not explicitly expressed in these models.

At the same time, there are many systems, in which it is unfeasible to introduce and preserve global time. For instance, it is proved that clock synchronization becomes impossible under definite conditions (Lamport, 1984; Dolev, et al, 1986; Fischer, et al, 1985; Attiya and Ellen, 2014).This precludes introduction of global time. As a result, in some systems, only local time (local time scale) can be treated in a consistent way. In addition, there are systems, in which global time (global time scale) coexists with a variety of local times (local time scales). Moreover, often these different times and time scales cannot be synchronized. All systems with these properties, which we call concurrent systems, cannot be portrayed by conventional models in general systems theory.

The goal of this paper is to construct more advanced than utilized now models of multi-agent distributed systems using the concept of local time, which exists and can be different in distinct components and parts of real systems according to the system theory of time (Burgin, 1992; 2002). These models provide descriptions and tools for exploration not only of classical systems with one global time but also of relativistic and concurrent systems, which can multiplicities of time.

It is interesting that absence of global time results in nonexistence of global states in a multicomponent system due to the concurrent functioning of the components and parts. As a result, time becomes multidimensional and demands specific unconventional mathematical structures for its representation.

In addition, exploring and modeling systems with a variety of independent and incoherent local times (local time scales), we come to the concepts of an observer, observation, synchronization and coordination of times and actions, which have not been studied in general systems theory. For instance, there are systems, in which it is possible to synchronize local time in different components and establish in such a way, global (absolute) time and a global time scale. However, conventional models from general systems theory give only a partial picture of these systems and do not allow exploration of synchronization. Note that synchronization plays a pivotal role in diversity of systems from organisms of people (cf., for example, (Winfree, 1987)) to computer networks (cf., for example, (Mills, 1991; Burgin, et al, 2016)) to distributed databases (cf., for example, (Lindsay, et al, 1979)) to physics and metrology (cf., for example, (Boixo, et al, 2006)) to human-computer interaction (cf., for example, (Burgin, et al, 2001)) to service systems (cf., for example, (Marzullo, 1983)). As Birman (2005) writes, clock synchronization is a necessary and critical part in most distributed systems.

To develop our model employing different structures of local time and local time scales, we use methods and approaches developed for concurrent processes in computer science and information theory. On the first stage of our research, we construct a kinetic system model by fundamentally advancing and further developing the Actor Model originally constructed for distributed computations (Hewitt, et al, 1973). Here we expand the scope of this model from computational systems and processes to general systems making it applicable for any system comprised of interacting subsystems, e.g., for organization, society, group of people or a computer network.

We call the basic component of the System Actor Model (SAM) constructed in this paper an *actor* although the conventional research typically uses the term *agent*. The reason for this is that according to the common usage, an agent is a system (an actor) who/that acts on behalf of another system (actor). Besides, in political science and economics, an agent is a person or entity able to make decisions and take actions on behalf of, or that impact, another person or entity called the principal (Rees, 1985; Eisenhardt, 1989).

This methodology allows treating agents as specific actors, who/which act on behalf other actors - principals. As a result, it is possible to represent any multi-agent multicomponent system by the System Actor Model but not every actor system can be represented by a multi-agent multicomponent system. There are many situations, especially, in society, when this difference between agents and free actors is very important. Taking into account that actors can be software systems, we see that software agents are a very special but essentially important case of actors. It is possible to compare relation between actors and agents with the relation between a function and a computable, e.g., recursive, function.

The System Actor Model is more flexible than agent models. For instance, agents are usually treated as autonomous systems perceiving with sensors and acting with actuators (Buşoniu, et al, 2010; Vlassis, 2007; Weiss, 1999). At the same time, actors can be directed or controlled by other actors. Some of actors can be without sensors and/or actuators. For instance, in problems of resource management, identifying each resource with an actor can make available a helpful, detailed perspective on the system while each of them might not have sensors and/or actuators and could be controlled and managed by a central authority.

This paper is structured in the following way. In Section 2, which goes after Introduction, we describe and explore the computational actor model. In Section 3, we construct and explore the system actor model, for which the computational actor model becomes a very special case. Besides, we go much further in comparison with the computational actor model by elaborating mathematical models of actors and environments where these actors function. This allows us to obtain many properties of actions, events, actors and their systems by rigorous mathematical techniques. One of the main targets of this work is to construct mathematical tools for exploration of collaborative and multi-agent systems, social network analysis and developing computational qualitative methods for data mining and digital humanities.

## 2. Actor model in computer science

The Computational Actor Model (CAM) and its methodology were developed in the theory of computation to provide constructive and theoretical tool for modeling, analyzing and organizing concurrent digital computations (Hewitt, et al, 1973; Hewitt, 2012). In CAM, actors are interpreted as computing devices or computational processes. We will call them *computational actors*.

To make the model uniform, the concept of a *messenger*, which is also a computational actor, is used instead of the concept of a *message*. An arbitrary event in the model is the receipt of a messenger, which impersonates a message, by the target (recipient) computational actor.

In CAM, computational actors perform computations based on information about other computational actors and asynchronously communicate using their addresses for sending and receiving messages. Additionally, computational actors can make decisions about their actions and behavior, create other computational actors, and determine how to react to the received messages. It is possible to treat all these actions as events in the space of computational actors although this not done in the original Computational Actor Model described in (Hewitt, et al, 1973).

Computational actors are described by two groups of axioms - structural axioms and operational axioms.

*Structural axioms* determine that the local storage of a computational actor may contain addresses of other computational actors such that satisfy one of the following conditions:

1. The addresses were provided when the computational actor was created.
2. The addresses have been received in messages.
3. The addresses were installed in computational actors created by the given computational actor.

*Operational axioms* determine what a computational actor can do. Namely, a computational actor can:

1. Create more computational actors
2. Send messages to other computational actors
3. Receive messages from other computational actors

Hewitt explains that CAM is rooted in physics while other theoretical models of computation are based on mathematics and/or logic (Hewitt, 2007). As a result, CAM has many properties similar to properties of physical models, especially, in quantum physics and relativistic physics. For instance, detailed observation of the arrival order of the messages for a computational actor affects the results of actor's behavior and can even increase indeterminacy. According to CAM, the performance of a computational actor is exactly defined when it receives a message while at other times, it may be indeterminate. Note that in reality, existence of nondeterministic models of computation, such as nondeterministic Turing machines (cf., for example, (Burgin, 2005)), shows that in some cases, the performance of a computing system or process cannot be exactly defined.

An important feature of CAM is that it can model systems that cannot be represented by the deterministic Turing's model while the latter is a special case of CAM. As Milner wrote (Milner, 1993), Hewitt had explained that a value, an operator on values, and a process could all be computational actors. Taking into account that computational actors can be interpreted as software systems, we can see that software agents are a very special but essentially important case of computational actors. The relation between computational actors and software agents is similar to the relation between the concept of a number and the concept of a rational number. As we know, there are numbers that are not rational.

Being very useful for concurrent computations, CAM has very limited applications beyond computer science. That is why, taking the concept of an actor in all its generality and building a mathematical representation of a system actor, for which a computational actor is a very particular case, we extend CAM far beyond the area of computers, computer networks and computations.

**3. Actor model in systems theory**

The basic concept in the System Actor Model (SAM) is the concept of an actor or more exactly, of a *system actor*, which, in particular, can be a computational actor. In what follows, we mostly call system actors simply by the name *actor* when it does not cause ambiguity.

Informally, a system actor is a system that functions in some environment interacting with other systems. It means that System Actor Model developed in this paper is inherently dynamic.

This notion of an actor is more formally described in the following way.

**Definition 3.1.** Taking a system $E$ of interacting systems $\{R_k \, ; \, k \in K\}$, which have the lower rank than $E$, we call the systems $R_k$ *actors* and treat them as actors in $E$, while $E$ is called the *environment* of each of the actors $R_k$.

Note that in contrast to the Computational Actor Model where computational actors are processes or operators, a system actor can be (or can be interpreted as) an arbitrary system or an element/component of an arbitrary system, e.g., people, social networks, living beings, cells of living beings, molecules, artificial systems, such as computers or computer networks, processes and/or imaginary systems, such as heroes of novels or movies. Besides, computational actors can perform only three types of actions – create new actors, send messages and receive messages (Hewitt, 2007). In comparison with these limited abilities, system actors, in general, can perform any actions. Possible actions are described by the axioms that determine the environment of system actors.

Although some authors call such systems by the name *agent* (cf., for example, (Doyle, 1983; Minsky, 1986)), it is more reasonable to call them actors because according to the common usage, an agent is a system (an actor) who/that acts on behalf of another system (actor). In addition, in political science and economics, an agent is a person or entity able to make decisions and take actions on behalf of another person or entity called the principal (Rees, 1985; Eisenhardt, 1989).

To build a mathematical model of an environment with actors, we construct a mathematical model (description) of an actor and an environment. Note that there is no similar mathematical model (description) in the computational Actor Model.

A *formal actor* (*system actor representation*) $A$ is described by a name and five structural components - three sets called *set components* of the actor $A$ and two functions (or relations) called *functional components* of the actor $A$. Namely, we have the following structure

$$A = (Rel_A, Act_A, Trn_A; React_A, Proact_A)$$

$A$ is a name of the actor

Three sets (set components) are:

- $Rel_A$ is the set of properties and relations of the actor $A$ (usually only relations in the environment $E$ are considered)
- $Act_A$ is the set of possible actions of the actor $A$
- $Trn_A$ is the set of possible actions aimed at the actor $A$

Two functions (functional components), which are multivalued in the general case, are:

The *reaction function* (*reaction relation*) shows responses of the actor $A$ to actions on $A$

$$React_A: Trn_A \to Act_A$$

The *proaction function* (*proaction relation*) shows actions of the actor $A$ instigated by properties and relations of $A$

$$Proact_A: Rel_A \to Act_A$$

Reactions and proactions determine behavior of the actor.

It is possible to consider the following example of a tentative proaction.

**Example 3.1.** If an actor $B$ is a friend of an actor $A$, then $A$ is doing something good for $B$.

The next example shows a prescribed proactions.

**Example 3.2.** If an actor $B$ is a friend of an actor $A$, then $A$ always accepts messengers (massages) sent by $B$.

As an example of reactions, we can consider the following situation.

**Example 3.3.** The *action* aimed at an actor $A$ is an e-mail from an actor $B$.

The *reaction* of $A$ is the response to this e-mail.

Relations between an actor and data structures or knowledge structures, which may also be represented as actors, can represent the memory of the actor. Then self-actions can change this memory performing computation, making decisions and deliberating subsequent actions. Note that it is possible to represent relations by properties and properties by relations (Burgin, 1985; 1990).

Parts and elements of actor's components have their modalities described below.

First, in this description of an actor *A*, it is useful to make a distinction between *actualized parts* (*elements*) and *tentative parts* (*elements*) of actor's components. For instance, some relations of *A* exist while others are only possible. Then the former relations are actualized while the latter are tentative. In a similar way, some actions have been performed or/and are performed while others are only possible. Then the former actions are actualized while the latter are tentative.

Second, if an actor has a knowledge system, then it is useful to make a distinction between *acknowledged parts* (*elements*) and *implicit parts* (*elements*) of actor's components. For instance, an actor *A* can know about some of its relations and do not know about others. Then the former relations are acknowledged while the latter are implicit.

Usually the components of an actor satisfy some restrictions. For instance, if an actor *A* is an automaton that does not give any output, e.g., if *A* is an accepting finite automaton, then all action of *A* are self-actions. In a formal setting, restrictions are described by axioms.

Properties, relations and actions have various properties including temporal properties. For instance, a *singular action* is performed at one moment of time, while performance of a *regular action* always demands some interval of time. In the theory of computational automata, all actions are singular (Burgin, 2005).

An important relation in this model is *acquaintance*. Namely, each actor *A* has a list of names (addresses) of *forward acquaintances* FAcq(*A*) and a list of names (addresses) of *backward acquaintances* BAcq(*A*). These lists regulate communication of the actor *A*. Namely, the actor *A* can send messages (messengers) only to forward acquaintances from FAcq(*A*) and can receive messages (messengers) from only backward acquaintances from BAcq(*A*). In particular, an actor (a system) can get feedback only from its backward acquaintances and can send feedback only to its forward acquaintances.

This assumption is formalized by the following axioms.

Let SMes(*A*, *B*) denotes the action of sending a messenger (a message) by an actor *A* to an actor *B*, $\Rightarrow$ denotes implication, $\Diamond$ denotes modal value "possible" and $\neg\Diamond$ denotes modal value "impossible". For instance, $\Diamond$ SMes(*A*, *C*) means that the actor *A* can send messages to the actor *C*.

**Axiom SM**. a) $\forall$ *A*, *C* (*C* $\in$ FAcq(*A*) $\Rightarrow$ $\Diamond$ SMes(*A*, *C*)).

b) $\forall A, C\ (C \notin \text{FAcq}(A) \Rightarrow \neg \Diamond \text{SMes}(A, C))$.

Informally, Axiom SMa means that the actor $A$ can send messages (messengers) to any of its forward acquaintances. Axiom SMb means that the actor $A$ cannot send messages (messengers) to any actor that (who) is not its forward acquaintance.

**Proposition 3.1.** If Axiom SM is true, then

$$\forall A, C\ (C \in \text{FAcq}(A) \Leftrightarrow \Diamond \text{SMes}(A, C))$$

*Proof.* By Axiom SMa, we have

$$\forall A, C\ (C \in \text{FAcq}(A) \Rightarrow \Diamond \text{SMes}(A, C))$$

Thus, we have to prove only

$$\forall A, C\ (\Diamond \text{SMes}(A, C) \Rightarrow C \in \text{FAcq}(A))$$

Let us assume that the actor $A$ can send messages to some actor $C$, i.e., $\Diamond \text{SMes}(A, C)$, but $C$ does not belong to the forward acquaintances of $A$, i.e., $C \notin \text{FAcq}(A)$. However, by Axiom SMb, we have $\neg \Diamond \text{SMes}(A, C))$ and by principle of the Excluded Middle, our assumption is incorrect. Thus, we have

$$\forall A, C\ (\Diamond \text{SMes}(A, C) \Rightarrow C \in \text{FAcq}(A))$$

Proposition is proved.

Let $\text{RMes}(C, A)$ denotes the action of receiving a messenger (a message) by an actor $A$ from an actor $C$.

**Axiom RM.** a) $\forall A, C\ (C \in \text{BAcq}(A) \Rightarrow \Diamond \text{RMes}(C, A))$.

b) $\forall A, C\ (C \notin \text{BAcq}(A) \Rightarrow \neg \Diamond \text{RMes}(C, A))$.

Informally, Axiom RMa means that the actor $A$ can receive messages (messengers) from any of its backward acquaintances. Axiom RMb means that the actor $A$ cannot receive messages (messengers) from any actor that (who) is not its backward acquaintance.

**Proposition 3.2.** If Axiom RM is true, then

$$\forall A, C\ (C \in \text{BAcq}(A) \Leftrightarrow \Diamond \text{Mes}(C, A))$$

*Proof* is similar to the proof of Proposition 3.1.

Note that $C \in \text{FAcq}(A)$ does not always mean that $A \in \text{BAcq}(C)$. Indeed, it is possible that an actor $A$ can send messages to an actor $C$ but $C$ cannot receive messages from $A$.

The following axiom for the environment $E$ remedies this situation.

**Connectivity Axiom CA.** $\forall A, C \in E\ (C \in \text{FAcq}(A) \Leftrightarrow A \in \text{BAcq}(C))$.

Informally, it means that an actor $A$ can receive messages (messengers) from an actor $B$ if and only if $B$ can send messages (messengers) to $A$.

Acquaintances that belong to both lists FAcq($A$) and BAcq($A$) are called *friends*. We denote this set by

$$Fr(A) = FAcq(A) \cap BAcq(A)).$$

In many cases (but not always), lists FAcq($A$) and BAcq($A$) coincide. In this case, they also coincide with the list Fr($A$).

Let us assume that Axioms CA, SM and RM are true.

**Proposition 3.3.** $\forall A, B \ (B \in Fr(A) \Rightarrow A \in Fr(B))$

*Proof.* The formula $B \in Fr(A)$ means that $B \in FAcq(A)$ and $B \in BAcq(A)$. By Axiom CA, we have

$$A \in BAcq(A) \text{ and } A \in FAcq(A)$$

Thus, $A \in Fr(B)$.

Proposition is proved.

Proposition 3.3 allows proving the following result.

**Proposition 3.4.** If in the environment $E$, all acquaintances are friends, then $E$ satisfies Axiom CA.

In the process of actor functioning, the lists of acquaintances and friends can change.

There are five basic types of actor relations:

- *Inner relations* are relations between parts and elements of the actor $A$. For instance, if an actor $A$ is an organization, then relations between members of this organization are inner relations of $A$.

- *Internal* relations are relations between the actor $A$ and its parts and elements. For instance, if an actor $A$ is an organization, then the relation "a member $H$ of $A$ receives salary from $A$" is an internal relation of $A$.

- *Outer relations* are relations of the actor $A$ to other actors, their parts, elements and the environment. For instance, if actors $A$ and $B$ are organizations, then cooperation between $A$ and $B$ is an outer relation of $A$.

- *Intermediate* relations are relations of parts and elements of the actor $A$ to other actors, their parts, elements and the environment. For instance, if an actor $A$ is an organization, then any relation between a member *H* of *A* and an actor *K* who is not a member of *A* is an intermediate relation of *A*.

- *External relations* are relations of other actors, in which the actor *A* is included. For instance, if actors are companies, then "to be a supplier" is an external relation of *A* when *A* is a supplier for another company.

Note that it is possible to consider actions, reactions and proactions as relations. However, it is more efficient to treat these structures separately making emphasis on the functionality and dynamics.

According to the theory of autopoiesis developed by Maturana and Varela (1973), relations and properties play a crucial role for autopoietic systems, which can be described briefly as self-producing devices, or a self-generating systems with the ability to reproduce themselves recursively. Relations and properties of a system determine the structure of this system (Burgin, 2012). Indeed, autopoietic systems are structure-determined systems according to the principle of structural determinism, which states that the potential behavior of the system depends on its structure (Maturana, 1997). It means that all actions of actors representing autopoietic systems are functions of relations and properties.

Observing actions in the real world, we see that there are different types, classes, groups and kinds of actions. Let us consider some of them.

Temporal characteristics of actions determines three groups of reactions and proactions:

- *Sharp immediate reaction* (*proaction*) of *A* starts immediately after the beginning of the corresponding action on *A* (immediately after the property or relation becomes overt).
- *Reserved immediate reaction* (*proaction*) of *A* starts when the corresponding action on *A* ends (when the corresponding property or relation becomes comprehensible).
- *Delayed reaction* (*proaction*) of *A* is performed when some time passes after the corresponding action on *A* (when some time passes after the corresponding property or relation becomes comprehensible).

Definitions imply the following results.

**Proposition 3.5.** If an action *a* is not immediate, then *a* and any sharp immediate reaction to *a* are parallel in time.

**Proposition 3.6.** An action and a reserved immediate reaction to it are strictly sequential in time.

**Proposition 3.7.** An action and a delayed immediate reaction to it are sequential in time.

There are other temporal relations between separate actions and events.

**Definition 3.2.** a) *Temporal independence* of events (actions) $E_1$ and $E_2$ means autonomy of their occurrence, i.e., either $E_1$ can take place before $E_2$ or $E_2$ can take place before $E_1$ or they can take place at the same time.

b) Two events (actions) are *temporally dependent* if they are not are temporally independent.

For instance, events in two disconnected computing systems are temporally independent. Note that disconnectedness means that these computers are not connected not only to one another but also to another system, for example, to the Internet. However, temporal independence does not prohibit simultaneous occurrence or coincidence of actions and events.

**Proposition 3.8.** Temporal dependence is a transitive relation.

Another important concept is temporal incomparability.

**Definition 3.3.** a) *Temporal incomparability* of events (actions) $E_1$ and $E_2$ means that it is not known whether they happen at the same time or which of them happens before the other.

b) Two events (actions) are *temporally comparable* if they are not are temporally incomparable.

For instance, events in two disconnected computers, which are not observed by the same observer, are temporally incomparable.

**Proposition 3.9.** Temporal comparability is a transitive relation.

Temporal independence and incomparability are related to concurrency.

**Definition 3.4.** *Concurrency* of two or more events or actions means their temporal independence and/or temporal incomparability, or in other words, that time of their happening is independent and sometimes incomparable.

As temporal independence allows simultaneous occurrence or coincidence, the introduced concept of concurrency comprises other interpretations of this term.

Concurrency is intrinsically related to such properties of events and actions as parallelism and sequentiality.

**Definition 3.5.** Two or more events or actions are *parallel* if their time intervals intersect (moments of their occurring coincide when they have zero duration, i.e., they are *momentary*).

For instance, when people read and understand some text, these actions are usually parallel but not always strictly parallel.

Note that independence of events allows them to be parallel. It implies that some parallel events can also be concurrent.

**Proposition 3.10.** If a momentary event (action) $E_1$ is parallel to a momentary event (action) $E_2$ and the event (action) $E_2$ is parallel to a momentary event (action) $E_3$, then all three events (actions) are parallel.

If the events are not momentary, then this result is not always true. For instance, let us consider events $E_1$, $E_2$ and $E_3$ such that $E_1$ starts at time 0 and ends at time 3, $E_2$ starts at time 2 and ends at time 5, and $E_3$ starts at time 4 and ends at time 7. Then the event $E_1$ is parallel to the event $E_2$ and the event $E_2$ is parallel to the event $E_3$, but the event $E_1$ is not parallel to the event $E_3$.

However, for interval events (actions), i.e., events (actions) with interval duration, it is possible to prove a result similar to Proposition 3.9.

**Proposition 3.11.** If an interval event (action) $E_1$ is parallel to an interval event (action) $E_2$, the event (action) $E_2$ is parallel to an interval event (action) $E_3$ and the event (action) $E_1$ is parallel to the event $E_3$, then all three events (actions) are parallel.

However, if the events are neither interval nor momentary, then this result is not always true. For instance, let us consider events $E_1$, $E_2$ and $E_3$ such that $E_1$ starts at time 0 and ends at time 3, $E_2$ starts at time 2 and ends at time 5, and $E_3$ starts at time 0 and continues to time 1, then restarts at time 4 and ends at time 7. Then the event $E_1$ is parallel to the event $E_2$ and the event $E_2$ is parallel to the event $E_3$, the event $E_1$ is parallel to the event $E_3$ but all three events are not parallel.

**Definition 3.6.** Two or more events or actions are *strictly parallel* if their beginning and end coincide and they go (take place) in the same time.

For instance, when the user switches her computer on (the first event), the computer starts working (the second event, which is strictly parallel to the first event).

**Proposition 3.12.** If an event (action) $E_1$ is strictly parallel to an event (action) $E_2$ and the event (action) $E_2$ is strictly parallel to an event (action) $E_3$, then the event (action) $E_1$ is strictly parallel to the event (action) $E_3$.

**Remark 3.1.** For parallel events (actions), this result is not always true.

**Definition 3.7.** a) Two events or actions are *sequential* if one of them, say $E_2$, starts after the other, say $E_1$, ends.

b) In this case, the event (action) $E_2$ is called *subsequent* to the event (action) $E_1$.

For instance, reception of information is subsequent to sending this information but usually it is not strictly subsequent.

**Proposition 3.13.** The relation between events and actions *to be sequential* is transitive.

Another important relation between events and actions is *to be strictly sequential*.

**Definition 3.8.** a) Two events or actions are *strictly sequential* if one of them, say $E_2$, starts exactly at the moment the other, say $E_1$, ends.

b) In this case, the event (action) $E_2$ is called *strictly subsequent* to the event (action) $E_1$.

In the theory of finite automata, it is assumed that starting from the second transition, each transition of the automaton is strictly subsequent to the previous transition (Burgin, 2005).

**Proposition 3.14.** If an event (action) $E_1$ is strictly subsequent to an event (action) $E_2$ and the event (action) $E_2$ has positive duration and is strictly subsequent to an event (action) $E_3$, then event (action) $E_1$ is not strictly subsequent to the event (action) $E_3$.

There are also structural characteristics of actions. One of them is direction.

Direction of actions determines three groups of actions:

- An *external action* of an actor is directed at other actors (cf. Figure 2).
- An *internal action* or a *self-action* of an actor is directed at the same actor and usually results in self-transformation (cf. Figure 1).
- A *combined action* of an actor is directed both at other actors and at the same actor(cf. Figure 3).

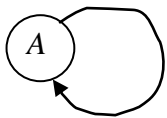

**Figure 1**. A *self-action* is an action of an agent on itself.

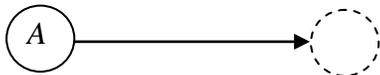

**Figure 2**. An *external action* is directed at other actors

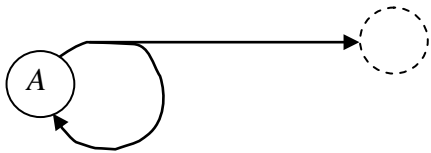

**Figure 3**. A *combined action* goes inside and outside.

**Example 3.4.** Reception of information is an example of a self-action.

**Example 3.5.** Computation performed by a system actor and any computational operation are examples of a self-action.

**Example 3.6.** Decision-making of a system actor is an example of a self-action.

**Example 3.7.** Sending information is an example of an external action.

**Example 3.8.** Working an inductive Turing machine transforms the content of its working register and from time to time, sends this content to the output register (Burgin, 2005). The action of the machine when it is doing both operations at the same time is a combined action.

Another structural characteristic of actions is modality, which determines the status of actions in the environment. There are three modalities of actions – positive, negative and neutral – and each of them contains four classes.

Positive modalities of actions:
- Possible actions
- Tolerable actions
- Permitted actions
- Performed actions

Negative modalities of actions:
- Impossible actions
- Intolerable actions
- Prohibited actions
- Not performed (but possible/permitted) actions

Neutral modalities of actions:
- Unknown actions
- Unidentified actions
- Unspecified actions
- Indefinite actions

There are definite relations between modalities of actions.

**Proposition 3.15.** a) Any unknown action is unidentified.

b) Any unidentified action is unspecified.

c) Any performed action is possible.

d) Any unknown possible and permitted action is not performed.

Structural characteristics of actions show that there are *simple actions* and *compound actions*, which are compositions of other actions. Compositions of actions are constructed using operations with actions. For instance, performing one action after another gives us the sequential composition of these actions.

If an action $a$ is a composition of actions $a_1$, $a_2$, $a_3$, …, $a_n$, for example, $a = \omega(a_1, a_2, a_3, …, a_n)$ where $\omega$ is an *n*–ary operation with actions, then any action $a_i$ ($i = 1, 2, 3, …, n$) *is included* in or is a *part* of the action $a$. It is denoted by $a_i \subseteq a$.

Informally, the relation $b \subseteq a$ means that performance of the action $a$ includes performance of the action $b$.

**Proposition 3.16.** For any actions $a$, $b$ and $c$, relations $a \subseteq b$ and $b \subseteq c$ imply the relation $a \subseteq c$.

Indeed, as a composition of compositions of actions is a composition of actions, relations $a \subseteq b$ and $b \subseteq c$ imply the relation $a \subseteq c$.

It means that the relation "to be a part" or "to be included" is transitive.

Composition preserves direction of actions.

**Proposition 3.17.** A composition of internal (external or combined) actions of the same actor is an internal (external or combined) action.

Organization of actions determines three groups of actions:

- *Direct actions* does not include additional operations (actions)
- *Mediated actions* include additional operations (actions or processes), for example, such as computation, meditation, contemplation or actions of other actors
- *Void actions* or *inactions*

Not to perform an action is also an action. It is a *void action*. All other actions are *proper actions*.

It is possible to build the system Actor Model (SAM) with one void action or with different void actions. It is possible to give a more precise description of actor's behavior when SAM allows different void actions. In this case, we have the following definition.

**Definition 3.9.** Not to perform an action $a$ is the *inaction* $\neg a$.

For instance, when a person $A$ is standing near the river and doing nothing seeing a person $B$ is drowning, this is a negative void action. When the Allies did nothing to prevent Hitler from seizing Austria and a part of Czechoslovakia, it was also a negative void action.

At the same time, there are positive void actions. For instance, when a person does not steal, it is a positive void action.

The concept of inaction or non-action plays an important role in Taoism because one of its central principles is the Principle of non-action (*Wu wei* in Chinese). Wu wei from the Tao Te Ching literally means non-action or non-doing and is connected to the paradox *weiwuwei*: "action without action" (Kirkland, 2004; Klaus, 2009).

Let us consider some properties of void actions.

**Proposition 3.18.** $\neg\neg a = a$.

Informally, it means that when non-doing of action *a* is not performed, then action *a* is performed. In essence, this is a version of the Principle of Excluded Middle because the proof of Proposition 3.18 uses this Principle and it is possible to consider systems of actors for which this assertion is not true.

Common sense tells us that independently in what way you compose non-doing, it will always be non-doing. We formalize this impression in the following axiom.

**Emptiness Axiom EA.** If $a_1$, $a_2$, $a_3$, …, $a_n$ are actions and ω is an *n*–ary operation with actions, then

$$\omega(\neg a_1, \neg a_2, \neg a_3, …, \neg a_n) = \neg \omega(a_1, a_2, a_3, …, a_n)$$

Axiom EA implies the following result.

**Proposition 3.19.** A composition of inactions is an inaction.

However, in general, Axiom EA is not always valid and a composition of inactions can be a proper action. For instance, let us consider the binary composition L(*x*, *y*), which combines two actions inferring the third action when only three actions can be performed. To provide an example of this situation, we can take the situation when a person can only either run (action *a*) or walk (action *b*) or stand (action *c*). Then combining two inactions ¬*a* (not running) and ¬*b* (not walking), we have L(*a*, *b*) = *c*, which is a proper action.

**Proposition 3.20.** If $a \subseteq b$, then $\neg b \subseteq \neg a$.

Indeed, if an action *b* includes an action *a*, then the absence of *a* implies and thus, includes, the absence of *b*.

It is useful to consider the *total inaction* $T_{IA}$ when simply nothing is done.

**Proposition 3.21.** For any action *a*, we have $\neg a \subseteq T_{IA}$.

Definitions imply the following result.

**Proposition 3.22.** A composition of non-void (proper) actions is a mediated action.

There other important types of actions.

A *primitive action* is a direct action that depends only on the input actions of other actors in the case of reactions or only properties and relations in the case of proactions.

An *automatic action* is a direct action that depends both on actions of other actors and on properties/relations.

Note that an inaction also can be primitive or automatic.

**Proposition 3.23.** When an action $a$ is primitive (automatic), the inaction $\neg a$ is also primitive (automatic).

Automatic actions allow unification of reactions and proactions in one (multivalued in a general case) function of combined actions

$$\text{Combact}_A: \text{Trn}_A \times \text{Rel}_A \to \text{Act}_A$$

In this context, the function $\text{React}_A$ is a restriction of the function $\text{Combact}_A$ when the action on $A$ is void and the function $\text{Proact}_A$ is a restriction of the function $\text{Combact}_A$ when the property/relation is void. This gives us the following result.

**Proposition 3.24.** Any primitive action is an automatic action.

Different types of actions spawn different types of actors.

**Definition 3.10.** A *behaviorally primitive actor A* has only primitive actions.

For instance, finite automata with one state are behaviorally primitive actors because their actions depend only on the input.

**Definition 3.11.** A *behaviorally automatic actor A* has only automatic actions.

For instance, finite automata are behaviorally primitive actors because their actions depend on both the input and inner state.

Proposition 3.24 implies the following result.

**Corollary 3.1.** Any behaviorally primitive actor is a behaviorally automatic actor.

There are various relations between actors.

**Definition 3.12.** Two actors are *identical* if they have the same structural components.

For instance, in contemporary industry, identical copies of many devices, such as vehicles, planes, computers and cell phones, are produced. In the system Actor Model, all these copies are represented by identical actors.

**Lemma 3.1.** Identity is an equivalence relation in sets of actors.

It is possible to find identical actors in many areas. One of them is theory and technology of information processing. Thus, there are models of computational systems, which contain many (sometimes, infinite) identical computing elements. Examples are cellular automata, artificial neural networks and iterative arrays.

For instance, a cellular automaton is a system of identical finite automata called cells, which form a net and interact with one another. A cellular automaton is determined by the following parameters (Burgin, 2005):

1. The *space organization* of the cells. In the majority of cellular automata, cells organized in a simple rectangular grid (mostly it is a one-dimensional string of cells and a two- or three-dimensional grid of cells), but in some cases, other arrangements, such as a honeycomb or Fibonacci trees.
a. The *topology* of the cellular automaton is determined by the type of the cell neighborhood, which consists of other cells that interact with this cell. In a grid, these are normally the cells physically closest to the cell in question. For instance, if each cell has only two neighbors (right and left), it defines linear topology. Such cellular automata are called linear or one-dimensional. It is possible to consider linear automata with the neighborhood of some radius $r > 1$. When there are four cells (upper, below, right, and left), the *CA* has two-dimensional rectangular topology. Such cellular automata are called planar or two-dimensional.
2. The *dynamics* of a cellular automaton, which determines by what rules cells exchange information with each other.

Traditionally, only rectangular organization of the cells and their neighborhoods has been considered for cellular automata. Recently, researchers have begun studies of cellular automata in the hyperbolic plane or on a Fibonacci tree (Margenstern, 2002). It is proved that such automata are more efficient than traditional cellular automata in the Euclidean plane. This higher efficiency is a result of a better topology in cellular automata in the hyperbolic plane.

According to the system Actor Model, each element of a cellular automaton is an actor and its actions consist of computing and communicating operations.

Looking at computer technology, we see that from the perspective of a manufacturer, products, e.g., computers, of the same type are identical.

Another important relation between actors is dynamic equivalence.

**Definition 3.13.** Two actors are *dynamically equivalent* if they have the same action components.

When it is necessary to solve the same problem for different input data, it is possible to use equivalent actors to this in a parallel or concurrent mode. This is often done in multiprocessor computers where identical processors perform necessary computations.

**Lemma 3.2.** Dynamic equivalence is an equivalence relation in sets of actors.

Identity of actors is a stronger relation than dynamic equivalence.

**Lemma 3.3.** Identical actors are dynamically equivalent.

Dynamic equivalence determines similarities in actor's behavior.

**Proposition 3.25.** An actor without actions is dynamically equivalent to an actor that has only void actions.

**Proposition 3.26.** An actor $A$ dynamically equivalent to a behaviorally primitive actor $B$ is behaviorally primitive.

**Proposition 3.27.** An actor $A$ dynamically equivalent to a behaviorally automatic actor $B$ is behaviorally automatic.

Another important relation between actors is homology.

**Definition 3.14.** Two actors $A$ and $B$ are *homological* if all their corresponding structural components are isomorphic.

For instance, for homological actors $A$ and $B$, there are isomorphisms between $Rel_A$ and $Rel_B$, between $React_A$ and $React_B$, and between $Proact_A$ and $Proact_B$.

**Example 3.9.** Let us consider two deterministic finite automata $A$ and $B$. They have the same set of states and the same set of start and final states. The first has the alphabet $\{0, 1\}$ and the second the alphabet $\{a, b\}$. Besides, all transitions of $A$ produced by input 0 are the same as all transitions of $B$ produced by input $a$ and all transitions of $A$ produced by input 1 are the same as all transitions of $B$ produced by input $b$. Then these automata are homological actors.

**Lemma 3.4.** Homology is an equivalence relation in sets of actors.

Identity of actors is a stronger relation than homology.

**Lemma 3.5.** Identical actors are homological.

Let us assume that isomorphisms between $React_A$ and $React_B$ and between $Proact_A$ and $Proact_B$. preserves primitive actions. Then we have the following result.

**Proposition 3.28.** An actor *A* homological to a behaviorally primitive actor *B* is behaviorally primitive.

Let us assume that isomorphisms between React$_A$ and React$_B$ and between Proact$_A$ and Proact$_B$. preserves automatic actions. Then we have the following result.

**Proposition 3.29.** An actor *A* homological to a behaviorally automatic actor *B* is behaviorally automatic.

A weaker type of relations is dynamic homology

**Definition 3.15.** Two actors *A* and *B* are *dynamically homological* if all their corresponding action components are isomorphic.

**Lemma 3.6.** Dynamic homology is an equivalence relation in sets of actors.

Dynamic equivalence of actors is a stronger relation than dynamic homology.

**Lemma 3.7.** Dynamically equivalent actors are dynamically homological.

Let us assume that isomorphisms between React$_A$ and React$_B$ and between Proact$_A$ and Proact$_B$. preserves primitive actions. Then we have the following result.

**Proposition 3.30.** An actor *A* dynamically homological to a behaviorally primitive actor *B* is behaviorally primitive.

Let us assume that isomorphisms between React$_A$ and React$_B$ and between Proact$_A$ and Proact$_B$. preserves automatic actions. Then we have the following result.

**Proposition 3.31.** An actor *A* dynamically homological to a behaviorally automatic actor *B* is behaviorally automatic.

According to their structure, we discern four classes of actors:

- A *structurally prime actor A* does not have components or parts.
- A *structurally primitive actor A* does not have components or parts, which are also actors.
- A *structurally composite actor A* has parts and/or components.
- A *structurally compound actor A* has parts and/or components, which are also actors.

In the actor's structure elements are also treated as parts.

The scale of observation defines what actors are prime. Thus, to be a prime actor depends on the scale of observation/treatment. For instance, in the observation scale of society, people are primitive actors. At the same time, in the observation scale of biology, people are composite actors.

The scale of modeling defines what actors are primitive. Thus, to be a primitive actor depends on the scale of modeling /representation. For instance, in the modeling scale of society, it is natural to represent people as primitive actors. At the same time, in the modeling scale of biology, it is natural to represent people as compound actors.

It is possible to develop a scale (ranging) of actors and deal with parts and components of a actor in this scale. Namely, an actor $A$ that is a part/component of another actor $B$ has lower range than $B$.

The system (environment) $E$ can be a model of a real system $R$, which can be physical, mental or structural. The system $R$ is called a *modeled domain* of $E$. In general, one environment $E$ can model different domains.

Let us consider a modeled domain $R$ of an environment $E$.

**Proposition 3.32.** If $R$ is the modeled domain of environment $E$ and a subdomain $P$ of $R$ is a modeled domain of $D$, then there is an injection of the set of all actors from $D$ into the set of all actors from $E$.

It is possible to introduce the following axiom

**Modeling Axiom MA**. Any object in the modeled domain $R$ is modeled by an actor in $E$.

If Pythagoras asserted "Everything is a number," the Modeling Axiom states "Everything and everybody is an actor."

The computational Actor Model that satisfies the Modeling Axiom is called the universe of CAM (Agha, 1986).

Let us consider an actor $A$ with the inner structure $Q$.

**Proposition 3.33.** If the Modeling Axiom is valid for an environment $E$ and its modeled domain $R$, then:

(a) Any structurally primitive actor is structurally prime.

(b) Any structurally composite actor is structurally compound.

**Corollary 3.2.** If the Modeling Axiom is valid for an environment $E$ and its modeled domain $R$, then there are only structurally primitive and structurally compound actors in $E$.

**Definition 3.16.** A *primary actor A* is not a part or component of other actors.

According to their communication, we discern five classes of actors – closed, inactive, generative, undemanding and open actors.

**Definition 3.17.** A *closed actor A* does not send and receive messengers (messages).

The concept of a closed actor allows treating almost anything, for example, tables, chairs, mountains, rivers, words, sounds, etc. as actors.

**Definition 3.18.** An *inactive actor A* does not send messengers (messages).

For instance, a sleeping woman does not send messengers (messages). Another example of an inactive actor is a receptor such as an automaton, which accepts input but gives no output (Burgin, 2005).

Definitions imply the following result.

**Lemma 3.8.** Any closed actor *A* is inactive.

The dual concept to an inactive actor is a non-receptive actor.

**Definition 3.19.** A *non-receptive actor A* does not receive messengers (messages).

An example of a non-receptive actor is a generator, i.e., such as an automaton, which does not accept input but gives output (Burgin, 2005). Another example of a non-receptive actor is a black hole (Thorne, 1994; Davies, 1995).

Definitions imply the following results.

**Lemma 3.9.** Any closed actor *A* is non-receptive.

It means that the property "to be closed" is stronger than the property "to be non-receptive."

**Lemma 3.10.** A non-receptive and inactive actor *A* is closed.

Opposite to closed actor are open actors.

**Definition 3.20.** An *open actor A* sends and receives messengers (messages).

Definitions imply the following results.

**Lemma 3.11.** Any open actor *A* is active.

It means that the property "to be open" is stronger than the property "to be active."

**Lemma 3.12.** A receptive and active actor *A* is open.

It is possible to distinguish actor by messages they send.

**Definition 3.21.** An *undemanding actor A* does not send requesting messengers (requests).

Definitions imply the following results.

**Lemma 3.13.** Any inactive actor *A* is undemanding.

Lemmas 3.9 and 3.13 imply the following result.

**Corollary 3.3.** Any closed actor *A* is undemanding.

It is possible to develop a scale (ranging) of actors and deal with parts and components of a primary actor in this scale.

Because an actor is functioning in some environment, it is also practical to use an extended actor representation, which includes relevant characteristics of the environment.

An *extended actor representation* consists of two names, three sets and four functions (or relations)

$$(A, E) = (Rel_A, Act_A, Trn_A ; React_A, Proact_A, VReact_A, VProact_A)$$

$A$ is a name of the actor.

$C$ is a name of the actor's environment.

Three sets are:

- $Rel_A$ is the set of properties of $A$ and relations of $A$ to other actors and the environment
- $Act_A$ is the set of possible actions of $A$
- $Trn_A$ is the set of possible actions on $A$

Four functions (multivalued in the general case) are:

The *reaction function* shows responses of $A$ to actions on $A$

$$React_A: Trn_A \to Act_A$$

*Proactions* show actions on $A$ instigated by properties and relations of $A$

$$Proact_A: Rel_A \to Act_A$$

For instance, if $B$ is a friend of $A$, then $A$ is doing something good for $B$.

The *virtual reaction function* shows responses of $A$ to all possible actions

$$VReact_A: Actp_E \to Act_A$$

Here $Actp_E$ is the set of all possible actions in $E$.

The *virtual proaction function* shows actions on $A$ instigated by all properties and relations, which exist in $E$

$$VProact_A: Relp_C \to Act_A$$

Here $Relp_C$ is the set of all possible properties and relations in $E$.

Definitions imply the following results.

**Lemma 3.14.** $React_A$ is a restriction of $VReact_A$.

**Lemma 3.15.** $Proact_A$ is a restriction of $VProact_A$.

In the System Actor Model, we also have a mathematical model of an environment.

An *environment representation* is described by a name, two sets and two functions (or relations)

$$E = (Relp_E, Actp_E, Trn_E ; EReact_E, EProact_E)$$

$A$ is a name of the actor

Two sets are:

- $Relp_E$ is the set of all possible properties and relations in $E$.

- Actp$_E$ is the set of all possible actions in *E*.

Two functions (multivalued in the general case) are:

E*Reactions* show all possible responses to actions in *E*

$$\text{EReact}_E: \text{Trn}_E \to \text{Act}_E$$

E*Proactions* show all possible actions instigated by properties and relations in *E*

$$\text{EProact}_E: \text{Rel}_E \to \text{Act}_E$$

Note that the systems $R_k$ in the environment *E* can have different ranks. For instance, in society, actors include separate individuals, organizations, countries, and so on.

**Definition 3.22.** a) If an actor *A* is a proper subsystem of an actor *B*, then the *rank* of *A* is lower than the rank of *B*.

b) If actors *A* and *B* consist of elements of the same rank, then the *rank* of *A* is equal to the rank of *B*.

By definition, the environment *E* has the highest rank in the system Actor Model.

**Proposition 3.34.** Elements, parts and components of an actor *A* have lower rank than *A*.

## 4. Conclusion

We have built a mathematical model of multicomponent interactive systems, which is called the System Actor Model and based on the formal structure of actors functioning in a multifarious convoluted environment. Different properties of such systems represented by an environment with actors have been obtained. Actions and events are analyzed in this context and different classes of events and actions are explicated and studied. Actors are also classified according to their traits. In addition, we elaborated a mathematical model of the environment. One of the main targets of this work is to construct mathematical tools for exploration of social systems. To conclude, we formulate some open problems for the System Actor Model.

The first cluster of problems is related to actions.

**Problem 1.** Formalize and study results of actions.

**Problem 2.** Formalize and study consequences of actions.

**Problem 3.** Formalize and study causes of actions.

**Problem 4.** Formalize and study in more detail structural, temporal and spatial characteristics of actions.

The second cluster of problems is related to actors.

**Problem 5.** Formalize and study tasks of actors.

**Problem 6.** Formalize and study obligations of actors.

**Problem 7.** Formalize and study norms of actors.

**Problem 8.** Formalize and study values of actors.

The third cluster of problems is related to concepts of agents and oracles, which are connected to the concept of actors.

**Problem 9.** Formalize and study relations between agents and actors.

**Problem 10.** Formalize and study relations between oracles and actors.

**References**

1. Agha, G. A. (1986) *ACTORS: A Model of Concurrent Computation in Distributed Systems*, The MIT Press Series in Artificial Intelligence, The MIT Press, Cambridge, Massachusetts
2. Attiya H. and Ellen, F. Impossibility Results for Distributed Computing, Morgan & Claypool Publishers, May 1, 2014
3. Barwise, J. and Seligman, J. (1997) *Information Flow*: *The Logic of Distributed Systems*, Cambridge Tracts in Theoretical Computer Science 44, Cambridge University Press
4. Bergson, H. *Time and Free Will*: *An Essay on the Immediate Data of Consciousness*, George Allen and Unwin, London, 1910
5. Birman, K.P. (2005) Clock Synchronization and Synchronous Systems, in *Reliable Distributed Systems*, Springer New York, pp. 493-508
6. Boixo, S., Caves, C.M., Datta, A. and Shaji, A. (2006) On decoherence in quantum clock synchronization, *Laser Physics*, v. 16, issue 11, pp. 1525-1532
7. Burgin, M. (1985) Abstract theory of properties, in *Non-classical Logics*, Institute of Philosophy, Moscow, pp. 109-118    (in Russian)
8. Burgin, M. (1990) Abstract Theory of Properties and Sociological Scaling, in *Expert Evaluation in Sociological Studies*, Kiev, pp. 243-264            (in Russian)
9. Burgin, M. A System Approach to the Concept of Time, *Philosophical and Sociological Thought*. - 1992. – No. 8 (in Russian and Ukrainian)
10. Burgin M. Time as a Factor of Science Development, *Science and Science of Science*, 1997, No. 1/2, pp. 45-59
11. Burgin, M. *Elements of the System Theory of Time*, LANL, Preprint in Physics 0207055, 2002, 21 p.       (electronic edition:   http://arXiv.org)
12. Burgin, M. *Super-recursive Algorithms*, Springer, New York/Heidelberg/Berlin, 2005
13. Burgin, M. (2012) *Structural Reality*, Nova Science Publishers, New York

14. Burgin, M., Karplus, W. and Liu, D. The Problem of Time Scales in Computer Visualization, in "Computational Science", *Lecture Notes in Computer Science*, v. 2074, part II, 2001, pp.728-737

15. Burgin, M., Mikkilineni, R. and Morana, G. Intelligent organisation of semantic networks, DIME network architecture and grid automata, *International Journal of Embedded Systems*, v. 8, No. 4, 2016, pp. 352-366

16. Buşoniu, L., Babuška, R. and De Schutter, B. Multi-agent reinforcement learning: An overview, in *Innovations in Multi-Agent Systems and Applications* (D. Srinivasan and L.C. Jain, eds.), v. 310 of Studies in Computational Intelligence, Berlin, Germany: Springer, pp. 183–221, 2010.

17. Davies, P. *About Time*, Simon & Schuster, New York/London/Tokyo, 1995

18. Dolev, D., Halpern, J.Y. and Strong, H. R. On the Possibility and Impossibility of Achieving Clock Synchronization, *Journal of Computer and System Sciences*, v. 32, 230-250 (1986)

19. Einstein, A., Lorentz, H.A., Weil, H., and Minkowski, H. *The Principle of Relativity*, Dover, 1923

20. Fischer, M.J., Lynch, N.A. and Paterson, M.S. Impossibility of Distributed Consensus with One Faulty Process, Journal of the Association for Computing Machinery, v. 32, No. 2, April 1985, pp. 374-382

21. Hewitt, C. (2007) What is Commitment? Physical, Organizational, and Social, Lecture Notes in Artificial Intelligence (Vázquez-Salceda, J. and Pablo Noriega, P. EdsS), Springer

22. Hewitt, C. (2012) What is computation? Actor Model versus Turing's Model, in *A Computable Universe, Understanding Computation & Exploring Nature as Computation* (H. Zenil, Ed.) World Scientific Publishing Company/Imperial College Press

23. Hewitt, C., Bishop, P. and Steiger, R. (1973) A Universal Modular Actor Formalism for Artificial Intelligence, IJCAI'73 *Proceedings of the 3rd international joint conference on Artificial intelligence*, Morgan Kaufmann Publishers Inc., San Francisco, CA, pp. 235-245

24. Kirkland, R. *Taoism*: *The Enduring Tradition*, Routledge, London/New York, 2004

25. Klaus, H. *The Tao of Wisdom. Laozi – Daodejing*. Chinese-English-German. Hochschulverlag, Aachen, 2009

26. Lamport, L. Unsolved problems, solved problems, and non-problems in concurrency , in *Proceedings of the 3rd ACM Conference on Principles of Distributed Computing*, 1984, pp. l-l 1

27. Lindsay, B. G., Selinger, P. G., Galtieri, C., Gray, J. N., Lorie, R. A., Price, T. G., Putzolu, F., Traiger, I. L., and Wade, B. W. Notes on distributed databases. IBM Res. Rep. RJ2571, IBM Research Division, San Jose, Calif., 1979.

28. Marzullo, K. *Loosely-Coupled Distributed Services: A Distributed Time System*, Ph.D. dissertation, Stanford University, 1983

29. Maturana, H. (1997) Metadesign, in Articulos y Conferences "*Diez Años de Post-Racionalismo en Chile*", Instituto de Terapia Cognitiva Web, Santiago http://www.inteco.cl/articulos/006/doc_ing1.htm last visited 9/6/2017

30. Maturana, H. and Varela, F. (1998) *The Tree of Knowledge*: *The Biological Roots of Human Understanding*, Shambhala, Boston

31. Mills, D.L. (1991) Internet time synchronization: the Network Time Protocol. *IEEE Trans. Communications COM-39*, v. 10, pp. 1482-1493

32. Milner, R. (1993) Elements of Interaction, *Communications of the ACM*, v. 36 No. 1, pp. 78-89

33. Prigogine, I. *From being to Becoming: Time and Complexity in Physical Systems*, San Francisco, 1980

34. Shoham, Y. and Leyton-Brown, K. *Multiagent Systems*: *Algorithmic, Game Theoretic and Logical Foundations*, Cambridge University Press, Cambridge, 2008

35. Thorne, K.S. *Black Holes and Time Warps*, Norton, New York, 1994

36. Vlassis, N. *A Concise Introduction to Multiagent Systems and Distributed Artificial Intelligence*, Synthesis Lectures in Artificial Intelligence and Machine Learning, Morgan & Claypool Publishers, 2007

37. Weiss, G. (ed.) *Multiagent Systems*: *A Modern Approach to Distributed Artificial Intelligence*, MIT Press, New York/London, 1999

38. Wiener, N. *Cybernetics, or Control and Communication in the Animal and the Machine*, MIT Press and Wiley, New York/London, 1961

39. Winfree, A.T. *The Timing of Biological Clocks*, Scientific American Library, New York, 1987